\documentclass[10pt, twocolumn]{article}
\usepackage{times,bm}
\usepackage[affil-it,]{authblk}
\usepackage[pagebackref=false,colorlinks=true,linkcolor=acsblue,citecolor=red,urlcolor=acsblue]{hyperref}
\usepackage{geometry}
\usepackage[switch*, displaymath,pagewise, mathlines]{lineno}
\usepackage{graphicx}
\usepackage{cite}
\usepackage{amsmath}
\usepackage{amsfonts}
\usepackage{amssymb}
\usepackage{slashed}
\usepackage{subfloat}
\usepackage{slashed}
\usepackage{color}
\usepackage{float}
\usepackage{multirow,multicol}
\usepackage{tabulary}
\usepackage[flushleft]{threeparttable}
\usepackage{pgfplots,subfigure}
\usepackage{xcolor}
\numberwithin{equation}{section}
\usepackage{tikz}
\usepackage{ulem}
\usepackage{hyperref}
\usepackage{nameref}
\usepackage{comment}
\usepackage{xcolor}
\definecolor{acsblue}{RGB}{17,76,139}
\definecolor{shadecolor}{RGB}{255,241,204}

\usepgflibrary{arrows}
\usetikzlibrary{shapes.callouts}
\tikzset{
	level/.style   = { thick, },
	connect/.style = { dotted, red   },
	notice/.style  = { draw, rectangle callout, callout relative pointer={#1} },
	label/.style   = { text width=2cm }
}

\usepackage{letltxmacro}
\makeatletter
\let\oldr@@t\r@@t
\def\r@@t#1#2{%
	\setbox0=\hbox{$\oldr@@t#1{#2\,}$}\dimen0=\ht0
	\advance\dimen0-0.2\ht0
	\setbox2=\hbox{\vrule height\ht0 depth -\dimen0}%
	{\box0\lower0.4pt\box2}}
\LetLtxMacro{\oldsqrt}{\sqrt}
\renewcommand*{\sqrt}[2][\ ]{\oldsqrt[#1]{#2}}
\makeatother
\usepackage{environ}

\begin{document}

\def\nofundrefquery{}
\def\stmdocstextcolor#1{}
\def\stmdocscolor#1{}
\newcommand{{\ri}}{{\rm{i}}}
\newcommand{{\Psibar}}{{\bar{\Psi}}}
\renewcommand{\rmdefault}{ptm}

\title{\mdseries{Rotational influence on fermions within negative curvature wormholes}}

\author{ \textit {\mdseries{Abdullah Guvendi}}$^{\ 1}$\footnote{\textit{ E-mail: abdullah.guvendi@erzurum.edu.tr } }~,~ \textit {\mdseries{Semra Gurtas Dogan}}$^{\ 2}$\footnote{\textit{ E-mail: semragurtasdogan@hakkari.edu.tr (Corr. Auth.)} }~,~ \textit {\mdseries{R. L. L. Vitória}}$^{\ 3}$\footnote{\textit{ E-mail: ricardo-luis91@hotmail.com} }   \\
	\small \textit {$^{\ 1}$ Department of Basic Sciences, Erzurum Technical University, 25050, Erzurum, Türkiye}\\
	\small \textit {$^{\ 2}$ Department of Medical Imaging Techniques, Hakkari University, 30000, Hakkari, Türkiye}\\
	\small \textit {$^{\ 3}$ Faculdade de Física, Universidade Federal do Pará, Av. Augusto Corrêa, Guamá, Belém, PA 66075-110, Brazil}}
	
\date{}
\maketitle

\begin{abstract}

\textcolor{black}{In this research, we examine relativistic fermions within the rotating frame of negative curvature wormholes. Initially, as is typical in our context, we introduce the wormholes by embedding a curved surface into a higher dimensional flat Minkowski spacetime. Subsequently, we derive the spacetime metric that characterizes the rotating frame of these wormholes. We then investigate analytical solutions of the generalized Dirac equation within this framework. Through exploring a second-order non-perturbative wave equation, we seek exact solutions for fermions within the rotating frame of hyperbolic and elliptic wormholes, also known as negative curvature wormholes. Our analysis provides closed-form energy expressions, and we generalize our findings to Weyl fermions. By considering the impact of the rotating frame and curvature radius of wormholes, we discuss how these factors affect the evolution of fermionic fields, offering valuable insights into their behavior.}

\end{abstract}

\begin{small}
\begin{center}
\end{center}
\end{small}

\bigskip


\section{\mdseries{Introduction}}\label{sec1}

Landau and Lifshitz introduced rotation to Minkowski spacetime, analyzing two specific effects: the impact on clocks within a rotating body and the singular behavior at significant distances within a uniformly rotating frame \cite{landau}. Notably, the latter effect imposes a limitation on the radial coordinate. Within the domain of relativistic quantum theory, investigating the impact of a rotating frame involves understanding the principles of special relativity, where the quantized nature of fields and particle behaviors are crucial considerations. This exploration unveils intriguing and unconventional effects absent in classical Sagnac-type descriptions, representing the coupling between angular momentum and rotation angular frequency of a uniformly rotating frame, emphasized across various references \cite{ref1, ref2, ref3, ref4}. In the non-relativistic context, this effect is known as the Page-Werner et al. term \cite{rot, ref12, rot2}. The incorporation of relativistic corrections within this framework underscores the potential for phase shifts induced by relative motion between an observer and the rotating frame. Analyzing relativistic quantum systems within such a rotating reference frame unveils a realm of interesting results. Additionally, for quantum fields within curved spaces, the modified effective gravitational field, influenced by the centrifugal force, alters the energy levels of quantum entities, especially in confined or bound states \cite{ref5}. In this context, several studies explore the impact of nontrivial rotating spacetimes on quantum mechanical systems. These investigations encompass various scenarios such as scalar bosons within cosmic string spacetime \cite{rot3}, a scalar field within a Kaluza-Klein theory \cite{rot4}, and a scalar field within time-dislocation spacetime \cite{rot5}. Predicting the effects of a rotating frame on relativistic quantum particles is theoretically plausible; however, practically observing these minute alterations demands precision, control, and advanced techniques in quantum optics, atomic physics, or condensed matter physics, alongside cutting-edge technology \cite{exp1, exp2, exp3}. Directly observing these effects necessitates trapping and controlling particles within the rotating frame, potentially detectable through variations in energy profiles, trajectories, or alterations in total angular momentum or spin. While extensive theoretical studies have scrutinized the effects of rotating reference frames on physical systems within non-relativistic and relativistic frameworks \cite{ricardo-1, ricardo-2, ref1, ref2, ref3, ref4, ref5, ref6, ref7, ref8, ref9, ref10, ref12, ref13, ref14, ref15, ref19, cjp}, there is a noticeable gap in announced results regarding the evolution of relativistic fermions within the rotating frame of graphene or a graphene wormhole characterized by a two-dimensional curved surface of constant negative curvature. This unexplored domain requires further investigation to comprehend the intricate interplay between relativistic fermionic fields and specific curved geometries such as graphene structures.

\textcolor{black}{In astrophysics, wormholes emerge from solutions in general relativity, serving as connections between two distinct regions \cite{einstein-rosen, wheeler, ellis}. These phenomena may manifest in both astrophysical contexts and condensed matter systems (by embedding curved surfaces into higher-dimensional spacetime (see also \cite{ES1,ES2})). However, the latter depiction differs because it involves two-dimensional curved surfaces, with time remaining unaffected by the background curvature \cite{wormhole}. Wormholes are actively researched to understand the fundamental properties of spacetime and whether they can exist within our current understanding of physics. Current research on wormholes spans from astrophysics to condensed matter contexts \cite{ap1, ap2, ap3, ap4, ap5, ap6, ap7, ap8, ap9, ap10, ap11, ap12, ap13, ap14}. Research examining quantum systems evolving within curved spaces has garnered considerable interest \cite{wormhole, wormhole2, Gibbons, 2, 2a, 2b, new1, new2, new3, new4, new5}, exemplified by investigations into Weyl particles in optical backgrounds \cite{2}, fermion-antifermion pairs in magnetized elliptic wormholes \cite{2b}, and more \cite{22c, 22d}. No results were found on fermionic fields in the rotating frame of negative curvature wormholes. This research addresses a gap in the literature by offering non-perturbative insights into fermions within rotating graphene and wormhole-like graphene structures characterized by a rotating surface with negative Gaussian curvature. Pursuing exact solutions holds promise in enabling the control or observation of compelling physical properties, potentially fostering novel phenomena, validating fundamental theories, and facilitating technological advancements. Our focus is on investigating relativistic fermion evolution within the rotating framework of negative curvature wormholes by seeking exact solutions to the corresponding Dirac equation.}

Our examination of Dirac particles within the rotating frame of negative curvature wormholes is structured as follows: In section \ref{sec2}, we unveil a general and non-perturbative wave equation, exploring the intricate realm of relativistic fermions in the rotating frame of negative curvature wormholes. Through a thorough analysis, we present the main wave equation governing their behavior in these environments. In section \ref{sec3}, we uncover precise and exact outcomes concerning the evolution of relativistic fermions within two distinct rotating frames: the hyperbolic wormhole and the elliptic wormhole. By investigating these specific scenarios, we reveal compelling insights into the behavior and dynamics of particles under these conditions. \textcolor{black}{Finally, in section \ref{sec4}}, we summarize our findings, encapsulating the essence of our results. Additionally, we provide a comprehensive discussion that explores the implications of our results across various physically plausible scenarios and limits. This reflective section aims to highlight the significance and broader ramifications of our work within the realm of theoretical physics.

\section{\mdseries{Wave equation}} \label{sec2}

\begin{figure}
\centering
\includegraphics[scale=0.60]{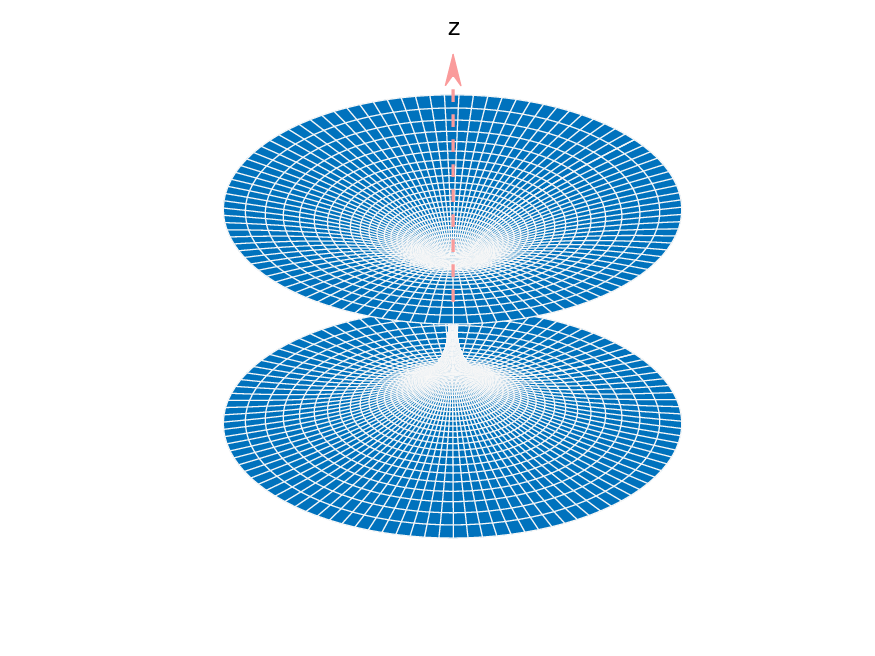}
\caption{\textcolor{black}{Geometric structure of the wormhole surface.}}
\label{fig:1}
\end{figure}

\textcolor{black}{In this section, we formulate a radial equation system for relativistic fermions in the rotating frame of negative curvature wormholes. This type of wormhole is characterized geometrically by points on its surface, which are expressed parametrically as shown below (also referenced in \cite{ES1,ES2}):
\[\vec{r}\left(u,v\right)=x\left(u,v\right)\hat{i}+y\left(u,v\right)\hat{j}+z\hat{k},\]
where \( x\left(u,v\right)=\chi\left( u \right)\cos(v) \), \( y\left(u,v\right)=\chi\left( u \right)\sin(v) \), and \( z\left(u\right)=\int {\sqrt{1-\chi_{,u}^2}}\,du \), with \( _{,u} \) indicating the derivative with respect to \( x^{u} \). The shape of the wormhole is generally described by the function \( \chi\left(u\right) \). The constraints on \( z \) arise from the relationship:
\[ds^2=dx^2+dy^2+dz^2=du^2+\chi^{2}\left( u \right)dv^2,\]
which results in the Hilbert horizon at \( \chi_{,u}=1 \) \cite{wormhole,Gibbons,wormhole2}. Embedding a three-dimensional wormhole into a (3+1)-dimensional spacetime generates effective gravity or curvature within the altered spacetime, leading to the experience of spacetime curvature by any particle or quasi-particle present in this modified background spacetime. Now, let us define the (2+1)-dimensional wormhole spacetime coordinates as \( dx^{\mu}= \{ cdT, du, dv \} \) and the (3+1)-dimensional Minkowski spacetime coordinates as \( dx^{\mu^{'}}= \{ cdT, dx, dy, dz \} \), where Greek indices denote the coordinates within the curved space. The transformation between these coordinates can be accomplished via the following matrix:
\[\frac{\partial x^{\mu^{'}}}{\partial x^{\nu}}=\left(\begin{array}{cccc}
1& 0 & 0& 0\\
0& \chi_{,u}\cos(v) & \chi_{,u}\sin(v) & \sqrt{1-\chi_{,u}^2}\\
0& -\chi_{,u}\sin(v) &\chi_{,u}\cos(v) &0
\end{array}\right).\]
Consequently, the metric of the considered wormhole spacetime takes the following form (with a negative signature) \cite{wormhole,2,2a,2b}:
\begin{flalign}
ds^2=c^2dT^2-du^2-\chi^2\left( u \right)dv^2.  \label{eq1}
\end{flalign}
A wormhole of this nature is geometrically depicted as shown in Fig. \ref{fig:1}. In Eq. (\ref{eq1}), \( c \) symbolizes the speed of light, and the function \( \chi(u) \) takes the form \( \chi\left( u \right)=a\, \cosh\left( \frac{u}{r_{0}} \right) \) for a hyperbolic wormhole and \( \chi\left( u \right)=b\ \sinh\left( \frac{u}{r_{0}} \right) \) for an elliptic wormhole. Here, \( a \) (\( b \)) denotes the radius of the associated wormhole at the midpoint (\( u=0 \)) between two ends, while \( r_{0}\) represents the radius of curvature of the wormhole surface \( \hat{u} \). It is noteworthy that the Gaussian curvature for the spacetime background as expressed in Eq. (\ref{eq1}) is \( K =-\frac{\chi_{,uu}}{\chi} \). Furthermore, Eq. (\ref{eq1}) describes a curved surface exhibiting negative Gaussian curvature, making such curved surfaces suitable for describing a monolayer graphene sheet and graphene wormholes. These surfaces can also be derived using the optical metric of the Banados-Teitelboim-Zanelli (BTZ) black hole \cite{Gibbons}, referred to as negative curvature wormholes (generally) or hyperbolic/elliptic wormholes depending on the shape of the radius function \( \chi\left( u \right) \) (for further details, see \cite{wormhole,Gibbons,wormhole2})}. In this scenario, we examine the rotating frame of spacetime given by Eq. (\ref{eq1}). This can be achieved through transformations provided by $T=t$, $u=\rho$, and $v=\phi+\omega_{rf} T$, where $\omega_{rf}$ denotes the angular frequency of the rotating frame. These transformations allow us to derive the spacetime interval describing a uniformly rotating frame of negative curvature wormholes (also see \cite{RF}).
\begin{flalign}
ds^2=(c^2-\omega_{rf}^2 \chi^2)dt^2-2\omega_{rf} \chi^2  dt d\phi-d\rho^2-\chi^2 d\phi^2,\label{eq2}
\end{flalign}
for which the Gaussian curvature remains unchanged but for $u \rightarrow \rho$. In this context, it is crucial to observe that the coordinate $\rho$ needs to adhere to a condition derived from the relationship $\chi\left(\rho\right)< c/\omega_{rf}$ for the fermions to reside within the light-cone. This introduces an additional requirement, indicating that the corresponding wave function must vanish at the point determined by the condition $\chi\left(\rho\right)= c/\omega_{rf}$. Referring to the line element Eq. (\ref{eq2}), one can derive the covariant $(g_{\mu\nu})$ and contravariant $(g^{\mu\nu})$ metric tensors
\begin{flalign*}
&g_{\mu\nu}=\left(\begin{array}{ccc}
c^2-\omega_{rf}^2 \chi^2& 0 & -\omega_{rf} \chi^2\\
0& -1 & 0\\
-\omega_{rf} \chi^2& 0 & - \chi^2
\end{array}\right), \nonumber\\
&g^{\mu\nu}=\left(\begin{array}{ccc}
1/c^2& 0 & -\omega_{rf}/c^2 \\
0& -1 & 0\\
-\omega_{rf}/c^2 & 0 & \omega_{rf}^2/c^2- \chi^{-2}
\end{array}\right).
\end{flalign*}
Now, let us express the Dirac equation in its generalized form within this specific background \cite{wormhole,2}
\begin{flalign}
&\left[\gamma^{\mu}\slashed{\mathcal{D}}_{\mu}+i\tilde{m}\mathcal{I}_{2}\right]\Psi(x^{\mu})=0, \quad \mu=t,\rho,\phi,\nonumber\\
&\slashed{\mathcal{D}}_{\mu}=\partial_{\mu}-\Gamma_{\mu},\quad \tilde{m}=\frac{mc}{\hbar}.\label{eq4}
\end{flalign}
In this context, $m$ denotes the fermion's rest mass, $\hbar$ represents the conventional Planck constant, $c$ stands for the speed of light, $\mathcal{I}_{2}$ signifies the 2-dimensional identity matrix, $\Psi(x^{\mu})$ symbolizes the Dirac field dependent on the spacetime position vector $x^{\mu}$, and $\gamma^\mu$ denotes the space-dependent Dirac matrices, derived from the relation: $\gamma^\mu=e^{\mu}_{k}\gamma^{k}$, where $e^{\mu}_{k}$ are the inverse tetrad fields ($k=0,1,2.$) and $\gamma^{k}$ denote the free Dirac matrices. The space-independent (free) Dirac matrices are represented in terms of the Pauli spin matrices ($\sigma^{x},\sigma^{y},\sigma^{z}$) as follows: $\gamma^{0}=\sigma^{z}, \gamma^{1}=i\sigma^{x}, \gamma^{2}=i\sigma^{y}$, in accordance with the signature of the line element \cite{new7,new8,new9}. In Equation (\ref{eq4}), $\Gamma_{\mu}$ refers to the spinorial affine connections for the Dirac field, attainable through the relation: $\Gamma_{\lambda}=\frac{1}{4}\left[e^{k}_{\nu_{,\lambda}}e^{\tau}_{k}-\Gamma_{\nu\lambda}^{\tau} \right]\mathcal{S}^{\mu\nu}$, where $_{,\lambda}$ indicates the derivative with respect to $x^{\lambda}$. Here, $\Gamma_{\nu\lambda}^{\tau}$ represents the Christoffel symbols $\Gamma_{\nu \lambda}^{\tau}=\frac{1}{2}g^{\tau \epsilon}\left[\partial_{\nu} g_{\lambda \tau}+\partial_{\lambda} g_{\epsilon \nu}-\partial_{\epsilon} g_{\nu\lambda} \right]$, $e^{\tau}_{k}$ denotes the inverse tetrad fields, and $\mathcal{S}^{\mu\nu}$ symbolizes the spin operators, defined as $\mathcal{S}^{\mu\nu}=\frac{1}{2}\left[\gamma^{\mu},\gamma^{\nu}\right]$ \cite{new7,new8,new9}. Now, the tetrads (and their inverses) can be obtained as follows (also refer to \cite{wormhole,2}):
\begin{flalign*}
e^{k}_{\mu}=\left(\begin{array}{ccc}
c& 0 & 0\\
0& 1 & 0\\
\omega_{rf} \chi & 0 & \chi
\end{array}\right), e^{\mu}_{k}=\left(\begin{array}{ccc}
1/c& 0 & 0\\
0& 1 & 0\\
-\omega_{rf}/c  & 0 & \chi^{-1}
\end{array}\right)
\end{flalign*}
since $g_{\mu\nu}=e^{k}_{\mu}e^{l}_{\nu}\eta_{kl}$ and $e^{\mu}_{k}=g^{\mu\nu}e^{l}_{\nu}\eta_{kl}$ where $\eta_{kl}$ is the Minkowski tensor, $\eta_{kl}=\textrm{diag}\left(1,-1,-1\right)$. Accordingly, we have $\gamma^{t}= \sigma^{z}/c, \gamma^{\rho}= i \sigma^{x}, \gamma^{\phi}=-\omega_{rf}/c\, \sigma^{z}+ i\,\chi^{-1}\sigma^{y}$ \cite{wormhole,2}. Also, non-vanishing components of the the spinorial affine connections are $\Gamma_{t}=\frac{i}{2}\omega_{rf} \chi_{,\rho}\sigma^{z}$ and $ \Gamma_{\phi}=\frac{i}{2}\chi_{,\rho}\sigma^{z}$. Hence, one can verify that \cite{wormhole,2}
\begin{flalign*}
\gamma^{\mu}\Gamma_{\mu}=-\frac{i}{2}\frac{\chi_{,\rho}}{\chi} \sigma^{x}.
\end{flalign*}
The metric Eq. (\ref{eq2}) allows us to factorize the Dirac field as the following $\Psi\left(x^{\mu}\right)=\textrm{e}^{-i\omega t}\textrm{e}^{is \phi}\left(\psi_{1}\left(\rho\right) \psi_{2}\left(\rho\right)\right)^{\textbf{T}}$ where $\omega$ is the relativistic frequency, $s$ is the spin and $\textbf{T}$ means transpose of the $\rho$-dependent spinor. Finally, a system of coupled equations can be derived
\begin{flalign}
&\left[\varpi-\tilde{m}\right]\psi_{1}-\left[\partial_{\rho}+\frac{1}{2}\frac{\chi_{,\rho}}{\chi}+\frac{s}{\chi}\right]\psi_{2}=0,\nonumber\\
&\left[\varpi+\tilde{m}\right]\psi_{2}+\left[\partial_{\rho}+\frac{1}{2}\frac{\chi_{,\rho}}{\chi}-\frac{s}{\chi}\right]\psi_{1}=0.\label{eq5}
\end{flalign}
in which, $\varpi=\omega/c+s  \omega_{rf}/c $, describing fermions within the rotating frame of negatively curved wormholes (see also \cite{Gibbons}). In the subsequent section, we aim to uncover analytical solutions to this wave equation within two specific contexts: (i) hyperbolic wormholes and (ii) elliptic wormholes.

\section{\mdseries{Exact results}} \label{sec3}

By decoupling the equations presented in Eq. (\ref{eq5}), a non-perturbative second-order wave equation for the component $\psi_{2}$ is obtained in the subsequent format
\begin{flalign}
\left[\partial_{\rho}^2+\frac{\chi_{,\rho}}{\chi}\partial_{\rho}
 +\hat{Q} +\left(\varpi^2-\tilde{m}^2\right)\right]\psi_{2}(\rho)=0,\label{eq6}
\end{flalign}
where
\begin{flalign}
\hat{Q}=\frac{1}{2}\frac{\chi_{,\rho \rho} }{\chi}-\frac{s^2}{\chi^2}-s\frac{\chi_{,\rho}}{\chi^2} -\frac{1}{4}\frac{\chi_{,\rho}^2}{\chi^2}.
\end{flalign}
Now, we have the ability to explore the precise solution of this wave equation under two distinct scenarios.

\subsection{\mdseries{Relativistic fermions in the rotating frame of hyperbolic wormhole}}\label{sec3.1}

If we examine a fermionic field within the relativistic framework of a rotating hyperbolic wormhole, the equation (\ref{eq6}) transforms into the following form
\begin{eqnarray*}
&\psi_{2_{,\rho\rho}}+\frac{\tanh\left( \frac{\rho}{r_{0}}\right)}{r_0}\,\psi_{2_{,\rho}}+\left[\tilde{\lambda}+\hat{Q}_{1}\right]\,\psi_{2}=0,\nonumber\\
&\hat{Q}_{1}=\frac{1}{2r_{0}^2}-\frac{s^2/a^2}{\cosh^2\left( \frac{\rho}{r_{0}}\right)}-\frac{s/a}{r_0}\,\frac{\tanh\left( \frac{\rho}{r_{0}}\right)}{\cosh\left( \frac{\rho}{r_{0}}\right)}-\frac{\tanh^2\left( \frac{\rho}{r_{0}}\right)}{4\,r^2_{0}},\label{eq7.1}
\end{eqnarray*}
where $\tilde{\lambda}=\varpi^2-\tilde{m}^2$. To transform this unfamiliar equation into a more recognizable form, we can employ an assumed function, $\psi_{2}\left(\rho\right)=(\sinh(\rho/r_{0})+i)^{p}(\sinh(\rho/r_{0})-i)^{p^{\textbf{*}}}\psi(\rho)$, where $p=\frac{3}{2}-i\frac{sr_{0}}{2a}$ and $^\textbf{*}$ denotes the complex conjugate. Then, by introducing a new variable, $\xi=\frac{1}{2}[1-i\ \sinh(\rho/r_{0})]$, we arrive at the following wave equation
\begin{flalign}
&\xi\left(\xi-1\right)\psi_{,\xi\xi}-[\gamma-(\alpha+\beta+1)\xi]\psi_{,\xi}+\alpha\beta\psi=0,\nonumber\\
&\alpha=1-ir_{0}\sqrt{\tilde{\lambda}}, \quad \beta=\alpha^{*}, \quad \gamma=\frac{3}{2}+i\frac{sr_{0}}{a}, \label{eq8}
\end{flalign}
in which $\alpha$, $\beta$ and $\gamma$ are constants. This equation is the renowned Gauss' hypergeometric equation, and its solution in the vicinity of the origin can be expressed as $\psi\left(\xi \right)=\mathcal{N}\  _{2}F_{1}\left(\alpha, \beta ; \gamma; \xi \right)$ \cite{Abramowitz}. Here, $\mathcal{N}$ denotes an arbitrary constant. The polynomial condition associated with this function is $\alpha=-n$ \cite{2,Abramowitz}, where $n$ signifies the principal quantum number $(n=0,1,2..)$. This criterion leads to the quantization principle governing the formation of the system under consideration, consequently yielding the subsequent expression
\begin{flalign}
\mathcal{E}_{ns}=-s  \hbar \omega_{rf} \pm \hbar c\sqrt{\frac{m^2c^2}{\hbar^2}-\frac{(n+1)^2}{r_{0}^{2}}}, \label{HW}
\end{flalign}
for the relativistic energy of the system under scrutiny. In this context, it becomes evident that the energy of such a system relies on the properties of particles, including spin (assuming $\omega_{rf}\neq0$), particle's rest mass, and the curvature radius of the hyperbolic wormhole, in addition to fundamental constants like $\hbar$ and $c$. The background under consideration is a spacetime with constant negative Gaussian curvature, characteristic of structures like monolayer graphene sheets or graphene wormholes since $K=-\frac{1}{r_{0}^{2}}$ (refer to \cite{Gibbons}). Consequently, there is potential utility in adjusting the outcomes for Weyl fermions by setting $m=0$ and substituting $c$ with $\nu_{f}$, where $\nu_{f}$ denotes the Fermi velocity, approximately equal to $c/300$. Notably, equation (\ref{HW}) transforms accordingly
\begin{flalign}
\mathcal{E}_{ns}=-s \hbar\omega_{rf} \pm i\hbar \nu_{f} \frac{(n+1)}{r_{0}},\label{HW-1}
\end{flalign}
if $m=0$. The Eq. (\ref{HW-1}) indicates that the energy of this system possesses a complex nature. Consequently, these states cannot remain stable, and the associated modes either decay or grow over time since $\Psi\propto \textrm{e}^{-i\frac{\mathcal{E}_{ns}}{\hbar}t}$. Additionally, the real oscillation of a Weyl particle is solely influenced by the coupling between the particle's spin and the rotational frequency of the frame. Our findings suggest that the stability of the background spacetime is compromised under perturbations caused by massless particles, while it remains stable under perturbations from Weyl antiparticles (also see \cite{2a,2b}) when $\omega_{rf}=0$. Notably, the coupling $s \omega_{rf}$ might induce symmetry breaking around the Dirac point when $m\neq 0$. Furthermore, it appears conceivable to measure, in principle, only the energy stemming from the $s\omega_{rf}$ coupling if the Dirac particle possesses a critical (or effective) mass value, $m_{c}=\frac{\hbar\ (n+1)}{c\ r_{0}}$. Such a scenario could arise particularly in condensed matter systems, especially when $r_{0}$ is on the order of a few nanometers (also see \cite{2,2a}).

\subsection{\mdseries{Relativistic fermions in the rotating frame of elliptic wormhole}}\label{sec3.2}

In this scenario, we derive precise outcomes for relativistic fermions within the rotating frame of an elliptical wormhole. Consequently, the wave equation presented in Eq. (\ref{eq6}) manifests as follows:
\begin{eqnarray*}
&\psi_{2_{,\rho\rho}}+\frac{\coth\left( \frac{\rho}{r_{0}}\right)}{r_0}\,\psi_{2_{,\rho}}+\left[\lambda+\hat{Q}_{2}\right]\,\psi_{2}=0,\nonumber\\
&\hat{Q}_{2}=\frac{1}{2r_{0}^2}-\frac{s^2/b^2}{\sinh^2\left( \frac{\rho}{r_{0}}\right)}-\frac{s/b}{r_0}\,\frac{\coth\left( \frac{\rho}{r_{0}}\right)}{\sinh\left( \frac{\rho}{r_{0}}\right)}-\frac{\coth^2\left( \frac{\rho}{r_{0}}\right)}{4\,r^{2}_{0}}.\label{eq8}
\end{eqnarray*}
Let us attempt to reformulate this equation using a new variable substitution, $z=\frac{1}{2}[1+\cosh(\rho/r_{0})]$, aiming to eliminate the hyperbolic functions. By expressing Eq. (\ref{eq8}) in terms of the variable $z$, we can render it free of hyperbolic functions, facilitating an exploration of its asymptotic behaviors. Consequently, employing an ansatz function, $\psi_{2}\left(z\right)=z^{\frac{2sr_0-b}{4b}}\left(z-1\right)^{-\frac{2sr_0+b}{4b}}\psi\left(z\right)$, the resulting equation can be transformed into the following wave equation
\begin{flalign}
&z\left(z-1\right)\psi_{,zz}-[\tilde{\gamma}-(\tilde{\alpha}+\tilde{\beta}+1)z]\psi_{,z}+\tilde{\alpha}\tilde{\beta}\psi=0,\quad \label{eq9}, \nonumber\\
&\tilde{\alpha}=-ir_{0}\sqrt{\tilde{\lambda}}, \quad \tilde{\beta}=\tilde{\alpha}^{*}, \quad \tilde{\gamma}=\frac{1}{2}+\frac{sr_{0}}{b}. \nonumber
\end{flalign}
In this context, $\tilde{\alpha}$, $\tilde{\beta}$, and $\tilde{\gamma}$ represent fundamental constants within Gauss hypergeometric wave equation, as extensively documented in references \cite{2,Abramowitz}. The regular solution of this equation in the vicinity of the origin can be elegantly expressed as $\psi\left(z \right)=\mathcal{C}\ _{2}F_{1}\left(\tilde{\alpha}, \tilde{\beta} ; \tilde{\gamma}; z \right)$, where $\mathcal{C}$ stands for an arbitrary constant. Remarkably, when the parameter $\tilde{\alpha}$ assumes the value $-n$, with $n$ representing the radial quantum number, the solution function $\psi(z)$ transforms into a polynomial of degree $n$ concerning the variable $z$. This establishes a link between the polynomial nature of the solution function and the quantum characteristics of the system, offering valuable insights into the system's energy behavior at different quantum states. This condition establishes an intriguing connection to the system's energy spectrum, unveiling the following set of energy levels
\begin{eqnarray}
\mathcal{E}_{ns}=-s  \hbar \omega_{rf} \pm \hbar c\sqrt{\frac{m^2c^2}{\hbar^2}-\frac{n^2}{r_{0}^{2}}}. \label{EW}
\end{eqnarray}
While bearing resemblance to the expression presented in Eq. (\ref{HW}), the relativistic energy pertaining to a fermion precisely equals the rest mass energy $\pm mc^2$ under specific conditions where $\omega_{rf}=0$ and $n=0$. This distinction stands as the fundamental dissimilarity between the outcomes delineated in Eqs. (\ref{HW}) and (\ref{EW}). Our investigations substantiate that within the rotating frame of an elliptic wormhole, the energy of a relativistic fermion manifests an intriguing independence from the curvature radius of the wormhole as the system converges toward its ground state. Notably, our observations assert that analogous deliberations, explicated in sec. \ref{sec3.1}, remain viable and generalizable when transitioning solely through $n+1\longrightarrow n$. The inference arises from the energy profiles derived in both situations, suggesting that the baseline energy level for relativistic fermions in the rotating frame of a hyperbolic wormhole appears akin to the first excited state for those in the rotating frame of an elliptic wormhole. Moreover, this correlation remains intact even when $\omega_{rf}=0$.

\section{Summary and discussions}\label{sec4}

In this study, we analyse the influence of non-inertial effects arising from the uniformly rotating reference frame of negative curvature wormholes, such as the hyperbolic wormhole and elliptic wormhole, on relativistic fermions. Our approach involves the investigation of analytically permissible solutions of the associated Dirac equation. Initially, we derive a comprehensive non-perturbative second-order wave equation. Subsequently, we proceed to ascertain the exact solutions of this wave equation within two distinct scenarios. In the first scenario, we focus on a background that characterizes the hyperbolic wormhole. Within this context, we determine the solution function expressed in terms of Gauss hypergeometric function. Consequently, we derive a non-perturbative energy spectrum in a closed mathematical form. The expression for this energy spectrum is provided by equation (\ref{HW}).

The findings reveal an intriguing interdependence between the energy states of relativistic fermions, the curvature radius of a wormhole, and the angular frequency of the uniformly rotating frame. Alongside the intrinsic quantum attributes of the fermionic field, our investigation underscores the direct influence of the particle's spin on the energy manifestation. Notably, this interaction may induce symmetry breaking centered around the Dirac point (zero energy), with energy levels diminishing until they reach a critical curvature radius of the wormhole, precisely at $r_{0}=\frac{\hbar(n+1)}{mc}$, where $n$ signifies the principle quantum number. Importantly, the system's stability is contingent upon the relationship between $r_{0}$ and $\frac{\hbar(n+1)}{mc}$; instability ensues if $r_{0}$ surpasses $\frac{\hbar(n+1)}{mc}$, irrespective of whether the angular frequency of the rotating frame is null or not. Our investigation further reveals intriguing insights into Weyl fermions ($m=0$), suggesting that states associated with these fermions may not exhibit stability due to potential damped modes characterized by a decay time $\tau_{n}=\hbar/|\mathcal{E}_{ns_{Im}}|$, where $\mathcal{E}_{ns_{Im}}$ denotes the imaginary component of the energy. In an alternative scenario involving an elliptic wormhole background in spacetime, our analysis of the primary equation yields precise solutions for the relativistic energy within the system under consideration. Surprisingly, the ground state energy identified in the earlier scenario appears to correspond to the first excited state of relativistic fermions within the rotating frame of an elliptic wormhole. Intriguingly, when the angular velocity of the rotating frame reaches zero, our results demonstrate that the relativistic energy of the system aligns with the rest mass energy of the fermion upon reaching the ground state ($n=0$). This alignment reflects an inevitable outcome given our understanding that the system will ultimately settle into its ground state. Furthermore, our outcomes suggest the possibility of measuring energy solely derived from the interplay between the particle's spin and the angular frequency of the uniformly rotating frame. This measurement becomes feasible if the particle possesses a critical (or effective) mass value represented by $m_{c}=\frac{\hbar (n+1)}{c r_{0}}$ and $m_{c}=\frac{\hbar n}{c r_{0}}$ for the first and second scenarios, respectively. Particularly intriguing is the potential manifestation of this specialized case in condensed matter systems, where $c$ can be equated to $\nu_{f}$ (Fermi velocity). Notably, this scenario appears plausible when the curvature radius of the wormhole approximates a few nanometers, assuming the particle's mass is the typical electron mass.

Recently, there has been a notable focus on performing thermodynamic evaluations using energy levels from quantum systems in complex environments. Several studies in the field have investigated this, setting the stage for future exploration of thermodynamic properties \cite{ter,ter1,ter2,ter4,ag1,ag2,ag3}. Our precise findings could be valuable for analyzing the thermal properties of graphene structures, presenting an intriguing avenue for future research.


\section*{\small{Data availability}}
This manuscript does not contain any associated data.

\section*{\small{Conflicts of interest statement}}

The authors have disclosed no conflicts of interest.

\section*{\small{Funding}}

This research has not received any funding.

\end{document}